\documentclass[sigconf]{acmart}

\usepackage{booktabs}
\usepackage{multicol}
\usepackage{graphicx}
\usepackage{caption}
\usepackage[T1]{fontenc}
\usepackage{balance}
\usepackage[utf8]{inputenc}
\usepackage{fixltx2e}
\usepackage{amsmath} 
\usepackage{dialogue}
\usepackage{pifont}
\usepackage{balance}
\usepackage[colorinlistoftodos,prependcaption,textsize=tiny]{todonotes}
\usepackage{tabularx, environ}
\newcommand{\cmark}{\ding{51}}%
\newcommand{\xmark}{\ding{55}}%

\def\BibTeX{{\rm B\kern-.05em{\sc i\kern-.025em b}\kern-.08emT\kern-.1667em\lower.7ex\hbox{E}\kern-.125emX}}

\setcopyright{acmlicensed}
\copyrightyear{2019} 
\acmYear{2019} 
\acmConference[CIKM '19]{The 28th ACM International Conference on Information and Knowledge Management}{November 3--7, 2019}{Beijing, China}
\acmBooktitle{The 28th ACM International Conference on Information and Knowledge Management (CIKM '19), November 3--7, 2019, Beijing, China}
\acmPrice{15.00}
\acmDOI{10.1145/3357384.3358047}
\acmISBN{978-1-4503-6976-3/19/11}
\begin{document}
\fancyhead{}
\title{Offline and Online Satisfaction Prediction in Open-Domain Conversational Systems}

\author{Jason Ingyu Choi}
\affiliation{%
  \institution{Computer Science Department\\ Emory University}
}
\email{in.gyu.choi@emory.edu }

\author{Ali Ahmadvand}
\affiliation{%
  \institution{Computer Science Department\\ Emory University}
}
\email{ali.ahmadvand@emory.edu}

\author{Eugene Agichtein}
\affiliation{%
  \institution{Computer Science Department\\ Emory University}
}
\email{eugene.agichtein@emory.edu}
\renewcommand{\shortauthors}{J. Choi et al.}

\begin{abstract}
Predicting user satisfaction in conversational systems has become critical, as spoken conversational assistants operate in increasingly complex domains. Online satisfaction prediction (i.e., predicting satisfaction of the user with the system after each turn) could be used as a new proxy for implicit user feedback, and offers promising opportunities to create more responsive and effective conversational agents, which adapt to the user's engagement with the agent. To accomplish this goal, we propose a conversational satisfaction prediction model specifically designed for open-domain spoken conversational agents, called ConvSAT. To operate robustly across domains, ConvSAT aggregates multiple representations of the conversation, namely the conversation history, utterance and response content, and system- and user-oriented behavioral signals. We first calibrate ConvSAT performance against state of the art methods on a standard dataset (Dialogue Breakdown Detection Challenge) in an online regime, and then evaluate ConvSAT on a large dataset of conversations with real users, collected as part of the Alexa Prize competition. Our experimental results show that ConvSAT significantly improves satisfaction prediction for both offline and online setting on both datasets, compared to the previously reported state-of-the-art approaches. The insights from our study can enable more intelligent conversational systems, which could adapt in real-time to the inferred user satisfaction and engagement.
\end{abstract}

\maketitle

\section{Introduction}
Conversational AI \citep{conver} has been an active area of research for decades, and based on recent advances in natural language understanding (NLU) and related fields, intelligent assistants have continuously improved. However, unlike human assistants, these bots do not understand the true meanings of their generated responses. They are more vulnerable to failures, and as of now, there have been no methods reported to reliably and automatically detect and correct failures in human-machine conversations as they occur. Introducing an accurate online satisfaction prediction model would spur dramatic improvements of conversational agents. For example, automatic and timely detection of failures would allow a conversational system to gracefully handle mistakes, and potentially improve both immediate and future system responses.

\begin{figure}[t!]
\includegraphics[height=225pt]{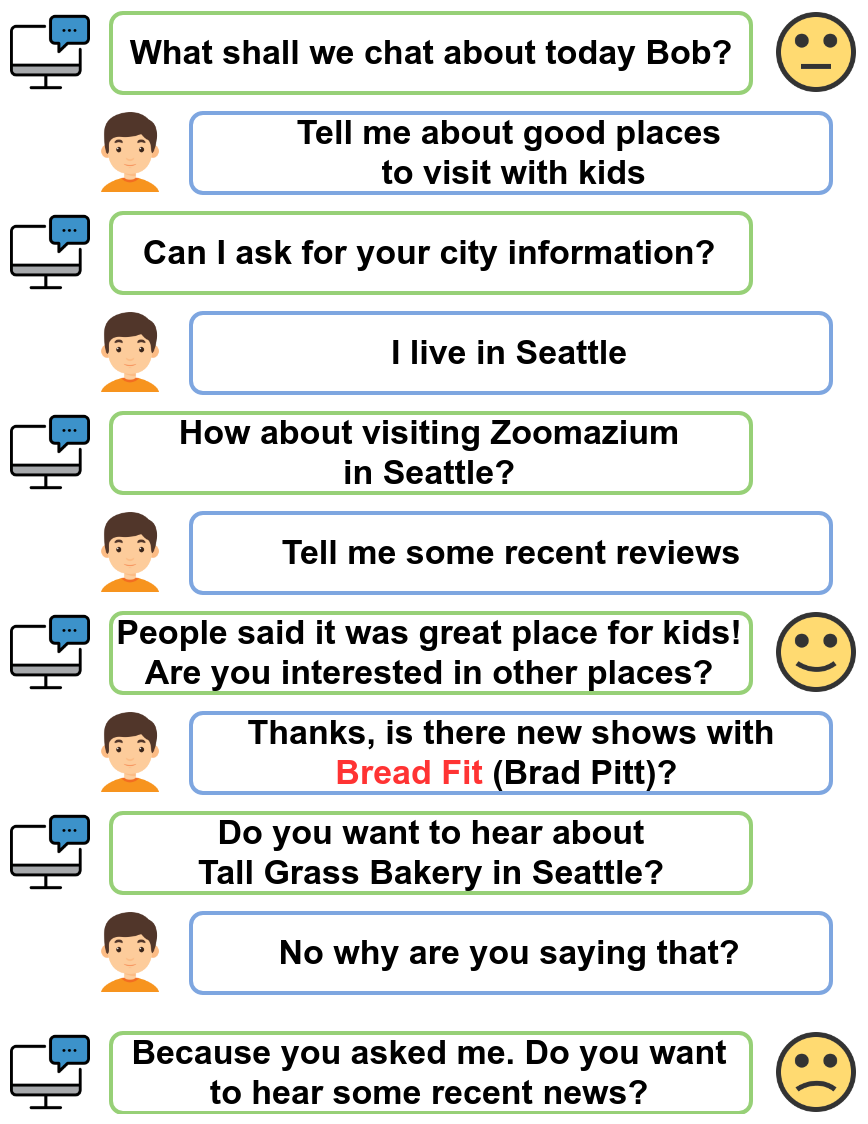}
\captionof{figure}{Sample human-machine conversation, with user satisfaction clearly decreasing as conversation progresses.}
\label{sample_conv}
\vspace{-6mm}
\end{figure}

Evaluating intelligent assistants is a challenging task, and has been an active area of research. For example, recent studies identified particular patterns of interactions which tend to contribute to final user satisfaction (e.g., \citep{kiseleva2016understanding, egregious, sensitive}), and new behavioral metrics such as conversational depth and topic diversity have been proposed to systematically evaluate user experience in conversational systems \citep{venkatesh2018evaluating, pred, conver}. However, as we will show, these metrics do not directly correspond to actual subjective and immediate user satisfaction with the conversation. Furthermore, predicting conversational satisfaction is significantly different from evaluating traditional informational systems, where signals such as clickthrough, dwell time, and transactional signals could be used to evaluate web search engines \citep{fox2005evaluating}, or touch-based features for satisfaction with mobile search and assistants \citep{pred, williams2016detecting}. Especially for open-domain conversational systems, since a user may not have clearly defined goals, a successful prediction system must understand a wide range of conversational intents, topic preferences, and user behavior signals.


As an illustration, consider a sample conversation of a user with our system, shown in the Figure \ref{sample_conv} above\footnote{Due to the Alexa Prize data confidentiality rules, we cannot reproduce actual user conversations, but the sample represents a typical conversation with our system}. The conversation starts well: our system successfully supports a 3-turn interaction about travel. However, our system failed to understand "Brad Pitt" due to automatic speech recognition (ASR) failure, and suggested a local bakery. The user has a hard time understanding the system's non-relevant response, and asks why we suggested bakery instead of movies. Our system lost context beyond this point, and suggested recent news, as a way of reclaiming the user's interest. At this point, the user is likely dissatisfied, as indeed supported by users' satisfaction ratings for such conversations.

Thus, evaluating immediate user satisfaction is crucial for developing adaptive conversational strategies such as failure recovery and topic switching. To address this problem, we propose a novel online conversational satisfaction prediction model (ConvSAT). ConvSAT first represents each conversation turn as a vector of carefully designed behavioral features, inspired by prior work (described next). These features aim to capture both the overall conversation state so far, and the immediate state of the conversation at the current turn. Being able to jointly model the overall (aggregated) conversation state and the immediate state after the current turn requires learning a complex interaction between the global and turn-level conversation state. We represent this interaction through a global/immediate feature matrix, and combine it with contextualized word encoders to learn conversation-specific contextual word representations conditioned on the conversation so far. Lastly, we enrich the word representations with sub-word (character) information, aiming to improve generalization on unseen and broken words from ASR output. 

Empirically validating such a satisfaction prediction system is challenging for two reasons. First, without a fully functional conversational agent, it is impossible to test and gather enough information to conduct a reasonable study. Second, even with an available system, it is challenging to create an environment and recruit enough users to interact with the system in a realistic setting. First, to calibrate our proposed method, we evaluate ConvSAT on a publicly available dialogue breakdown detection challenge 3 (DBDC3) dataset, generated by human users talking to different chatbots. The results show that our method outperforms the more recently reported state of the art methods on that task. Having established the effectiveness of our prediction method under controlled conditions, we then report satisfaction prediction results over a large conversational dataset collected by conversing with real users during the Amazon Alexa Prize 2018 competition. Our system successfully conversed with thousands of Alexa users, with a large fraction of these conversations explicitly rated by users for perceived conversation quality. Together, these results establish that our proposed method not only outperforms the existing state of the art methods on an established benchmark, but is able to successfully generalize to the more challenging real-world scenario of Alexa-based open-domain conversational AI challenge. In summary, our contributions include:
\begin{itemize} 
\item {A novel ConvSAT model, applicable to both conversational satisfaction and failure prediction tasks, operationalized in both online and offline settings.}
\item {A comprehensive behavioral feature matrix, designed to capture both immediate and aggregated evidence for conversational user satisfaction.}
\item{Extensive experiments, demonstrating the effectiveness of ConvSAT on an open benchmark dataset, and on real open-domain human-machine conversations.}
\end{itemize}

\begin{figure*}[h]
\centering
  \includegraphics[width = 460pt]{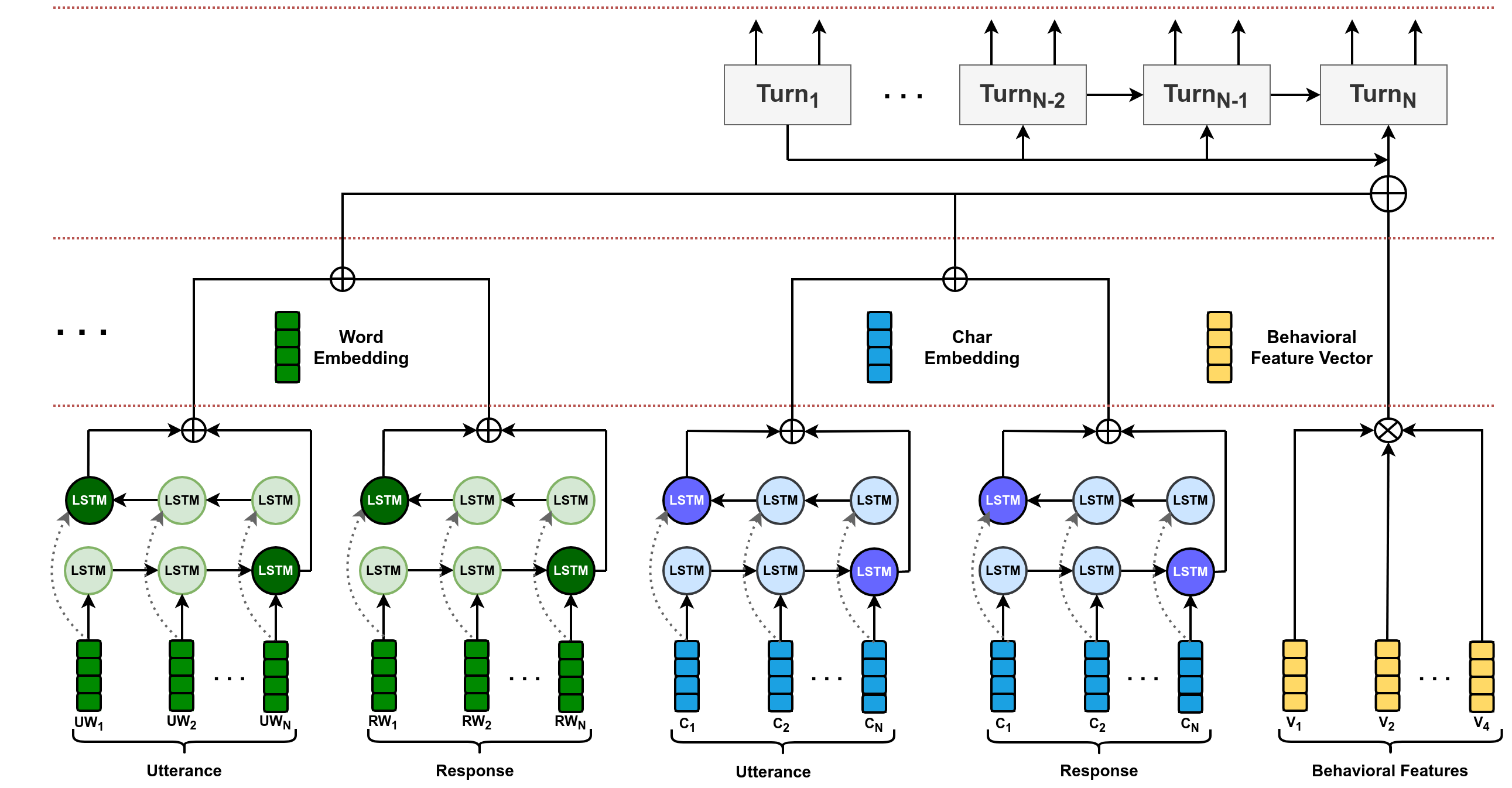}
\vspace{-2mm}
\captionof{figure}{Overview of the ConvSAT model  (best viewed in color).}
\label{ConvSAT}
\vspace{-3.5mm}
\end{figure*}

\vspace{-1mm}
\section{Related Work}
In this section, we summarize the related work on conversational AI and satisfaction prediction to put our contributions in context. 

\vspace{-1.5mm}
\paragraph{\bf Conversational AI}
Conversational systems \citep{conver} aim to interact with users naturally through conversations. One classic study summarized four main challenges of conversational systems as: 1) processing and understanding the noisy text from speech; 2) designing a flexible system for easy adaptation of different tasks; 3) domain and intent recognition; 4) mixed-initiative dialogue between machines and users \citep{toward_conv}. There has been dramatic recent progress in all of these areas, which enabled modern conversational systems. For example, ASR quality has improved drastically over the last few years \citep{speech_Hinton}, enabling more natural voice input. Both rule-based dialogue management (DM) systems \citep{form-DM, ravenclaw} and end-to-end DM systems using neural networks \citep{DM_end, luo2018learning, dhingra2016towards} have also improved in sophistication and flexibility \cite{alquist}. To enable an easier extension to new tasks, modular architectures with central DM systems have been proposed \citep{emory, gunrock, sounding_board}. 

To improve the knowledge retrieval process, many recent approaches introduced novel frameworks to incorporate external knowledge into response generation as well as actively learning concepts through conversations \citep{dinan2018wizard, luo2018learning, jia2017learning, ghazvininejad2018knowledge}. Incorporating artificial personalities are studied to improve empathetic, personalized engagements within these systems \citep{sounding_board, xiaoice}. 

Despite significant advances in conversational AI, a challenge of conducting a natural, mixed-initiative conversation remains elusive because deciding when to lead the conversation or to follow the user's interest heavily depends on each user. Especially for open-domain conversations that do not have clearly defined goals, this becomes more problematic. One active field related to this challenge is predicting or evaluating user satisfaction interactively with intelligent assistants. If a user's satisfaction or engagement with a conversation could be predicted in real-time, a conversational system could better adapt to the user's interests or intents, or initiate graceful failure correction, among many other possible actions. 

\vspace{-1.5mm}
\paragraph{\bf Measuring and Predicting User Satisfaction}
User satisfaction can be viewed as an attitude toward an information system, which is measured by various types of beliefs about user interactions as defined in \citep{user_satisfaction_theory1, user_satisfaction_theory2, paradise1, paradise2}. More precisely, recent works on evaluating intelligent assistants focus on extracting useful features and representations to train an offline satisfaction model. For traditional IR systems such as Web search engines, previous studies showed that incorporating implicit features such as deviations from the average behavior and time on page into the ranking function could improve the search results \citep{eugene_web}. For mobile search assistants, combining implicit features with additional touch-related features dramatically increased the performance of a trained satisfaction model \citep{pred, kiseleva2016understanding}. Thus, we hypothesize that extracting contextualized behavioral features could lead to a more robust conversational satisfaction model.

To represent context, we draw on active prior work on representation learning using unsupervised feature discovery to predict user satisfaction. One recent work proposed a query representation learning technique with intent-sensitive word embeddings, and showed that modifications to improve query representation can improve overall model performance \citep{sensitive}. Another recent work introduced a model that can detect egregious conversations using textual representations, and addressed how this technique can be applied to an automated evaluation scheme \citep{egregious}. There have been studies to predict causes of query reformulation in intelligent assistants by using system, acoustic, language and additional features \citep{sano2017predicting}. Hence, we incorporated query reformulation features such as query overlap and repetition counts in our study to capture query reformulation as a potentially negative signal.

Lastly, there have been efforts in restricted domains to predict online satisfaction signals, such as using manually curated features from a flight-booking system \citep{turn_prediction} or detecting online dialogue breakdowns (dissatisfaction) from DBDC3 challenge \citep{dbdc3, kth}. Another recent approach proposed a novel self-feeding framework to improve the quality of conversational systems \citep{hancock2019learning} using online predictions. We will extend the proposed ideas here by introducing a much more comprehensive set of features to predict online satisfaction in non-goal oriented conversations. In summary, this is the first report of predicting satisfaction of real users on real-world open-domain conversations, for both the offline and online settings. 

\section{C\lowercase{onv}SAT: Method Description}
\label{approach}
In this section, we present our proposed conversational satisfaction prediction model (ConvSAT). As illustrated in Figure \ref{ConvSAT}, ConvSAT considers three complementary input dimensions: 1) contextualized word encoders (colored green); 2) contextualized character encoders (colored blue); 3) behavioral feature matrices (colored yellow).

\vspace{-1.5mm}
\subsection{Model Architecture}
\paragraph{\bf Contextualized Word Encoders}
\label{word-encoders}

To add context history, we define a hyper-parameter called context window size (\textit{W}) to control how many previous turns to condition. To ensure an online setting, we do not incorporate any future information. Hence, given previous turns (\textit{T\textsubscript{1}} ... \textit{T\textsubscript{i-1}}) and current turn (\textit{T\textsubscript{i}}), current utterance (\textit{U\textsubscript{i}}) and current response (\textit{R\textsubscript{i}}) are expanded with previous \textit{W} turns (We fixed \textit{W}=3 for the illustration purpose throughout this section):

\vspace{-5mm}
\begin{gather}
\label{eq1}
U_{i} = [U_{i-3}; U_{i-2}; U_{i-1}; U_{i}]
\\
R_{i} = [R_{i-3}; R_{i-2}; R_{i-1}; R_{i}] 
\end{gather}
\vspace{-5mm}

The boundaries between the expanded utterances and responses are marked with special tokens <U-END> and <R-END>. These two expanded sequences are tokenized to obtain two word sequences (\textit{U\textsubscript{i}\textsuperscript{w}}, \textit{R\textsubscript{i}\textsuperscript{w}}), which will be the inputs to contextualized word encoders:

\vspace{-8mm}
\begin{gather}
\textit{U\textsubscript{i}\textsuperscript{w}} = [U_{w1}; U_{w2}; U_{w3}; U_{w4} \ ... \ U_{wn}] 
\\
\textit{R\textsubscript{i}\textsuperscript{w}} = [R_{w1}; R_{w2}; R_{w3}; R_{w4} \ ... \ R_{wn}]
\end{gather}

To represent the utterances contextually, we chose bidirectional Long Short Term Memory (bi-LSTM) networks, as they have shown promising performance for representing text. We have two separate encoders for both utterances (\textit{Encoder\textsubscript{U}}) and responses (\textit{Encoder\textsubscript{R}}). This is because in human-machine conversations, the ratio of words in an utterance to response is low, mainly due to limitations in open-domain conversational systems. By using two separate encoders, the goal is to reduce the possible bias towards long responses. The last hidden outputs from each forward LSTM (\( \overrightarrow{h_n})\) and backward LSTM (\( \overleftarrow{h_n})\) are concatenated to represent the entire word semantics in \textit{U\textsubscript{i}} and \textit{R\textsubscript{i}}. These two outputs are concatenated to obtain the final context representation (\textit{Encoder\textsubscript{word}}) at \textit{T\textsubscript{i}}:

\vspace{-3mm}
\begin{gather}
Encoder\textsubscript{word} = [Encoder_U;Encoder_R]
\end{gather}

\paragraph{\bf Contextualized Character Encoders}
Voice-based conversational systems are vulnerable to automated speech recognition (ASR) errors. Errors are more frequent for entity names, such as people or brand names, and transcription errors in these are likely to lead to a failed conversation. We noticed that mis-spelled or mis-segmented words often share similar sub-word structures, because various accents and pronunciations are originated from a single root word. As an illustration, consider a short example of how ASR recognized several automobile brands for people with foreign accents:

\vspace{1mm}
\begin{center}
    \begin{tabular}{l|l}
    \toprule
    \bf Actual word & \bf ASR failures \\
    \bottomrule
        Mercedes & Sadis, Cedes, Sadi's \\
        McLaren & Mac Laren, Mac Lauren, Mclaurin \\
        Aston Martin & Astone Martine, Ask Tony Martin \\
    \bottomrule
    \end{tabular}
\end{center}
\vspace{1mm}

Without subword (character-level) information, these errors are likely to create noise in learning robust word representations. Moreover, the frequency of errors such as \underline{Sadi's} appearing in our data is low, which causes the embedding matrix to be more sparse. For the \underline{Ask Tony Martin} case, it is likely that the model will understand this phrase differently from the original intent. Hence, by jointly training word-level and sub-word (character-level) models, we hypothesize that the overall semantics can be modeled better.

From the expanded word sequences \textit{U\textsubscript{i}\textsuperscript{w}}, \textit{R\textsubscript{i}\textsuperscript{w}} in (3) and (4), we derive the character sequences \textit{U\textsubscript{i}\textsuperscript{c}} and  \textit{R\textsubscript{i}\textsuperscript{c}}:

\vspace{-3mm}
\begin{gather}
U\textsubscript{i}\textsuperscript{c} = [[c_{1,1} \ ... \ c_{1,k}]; [c_{2,1} \ ... \ c_{2,k}] \ ... \ [c_{n,1} \ ... \ c_{n,k}]] 
\\
R\textsubscript{i}\textsuperscript{c} = [[c_{1,1} \ ... \ c_{1,k}]; [c_{2,1} \ ... \ c_{2,k}] \ ... \ [c_{n,1} \ ... \ c_{n,k}]] 
\end{gather}
\vspace{-3mm}

The following \textit{U\textsubscript{i}\textsuperscript{c}} and \textit{R\textsubscript{i}\textsuperscript{c}} are 2-dimensional matrices with first dimensions representing each tokenized word and second dimensions representing characters of each word. We flatten these matrices to two 1-dimensional character sequences. We also used bi-LSTM networks (Encoder\textsubscript{Uc}, Encoder\textsubscript{Rc}) to obtain final character representation (\textit{Encoder\textsubscript{char}}), which is identical to the process in (5).


\vspace{-1mm}
\paragraph{\bf Behavioral Features with Online Scaling}

Behavioral features are manually engineered to encode different aspects of user behavior. At a particular turn \textit{T\textsubscript{i}}, user behavior is represented as one feature vector (\textit{v\textsubscript{i}}), which can be a concatenation of various types of features. To incorporate conversational context, we append last \textit{W} feature vectors to obtain matrix \textit{V\textsubscript{i}}:

\vspace{-3mm}
\begin{gather}
V\textsubscript{i} = [v_{i-3}; v_{i-2}; v_{i-1}; v_{i}]
\end{gather}
\vspace{-3mm}

Each \textit{v\textsubscript{n}} encodes local information from beginning turn \textit{T\textsubscript{0}} to turn \textit{T\textsubscript{n}}. For instance, if we count total words in current \textit{T\textsubscript{i}}, total words are counted from \textit{T\textsubscript{0}} to \textit{T\textsubscript{i}}. Similarly, when computing the average number of words, total words from \textit{T\textsubscript{0}} to \textit{T\textsubscript{i}} is divided by the current turn \textit{i}. Our proposed scaling function \textit{S}(\textit{v}, \textit{i}) scales feature vectors (v) with respect to the current turn index (i). For online predictions, such scaling mechanism is crucial, because the goal is to detect a relative change in user behavior as the conversation progresses. If a user engaged deeply in one topic but started to diverge in the later turns, a feature capturing topic transition rate (how likely conversational states change) will gradually increase from lower to higher values. We apply this online scaling function to each vector in \textit{V\textsubscript{i}} to obtain scaled \( \hat{V\textsubscript{i}} \):

\vspace{-3mm}
\begin{gather}
\hat{V\textsubscript{i}} = [S(v_{i-3},i-3), \ S(v_{i-2},i-2) \ ... \ S(v_{i},i)]
\end{gather}
\vspace{-3mm}

The resulting \( \hat{V\textsubscript{i}} \) is a 2-dimensional dense matrix, with row representing each turn and column representing each scaled feature in respect to that turn \textit{i}. Then, we feed \( \hat{V\textsubscript{i}} \) to an attention layer to obtain a weighted sum of each vector. Given each \textit{v\textsubscript{i}}, similarity score s\textsubscript{i} is computed based on a shared trainable matrix \textit{M}, feature context vector \textit{c} and a bias term \textit{b\textsubscript{i}}. \textit{M}, \textit{c} and \textit{b\textsubscript{i}} are initialized randomly and jointly learned during training. Softmax activation is applied to similarity scores to obtain attention weights \( \alpha \). Lastly, using learned \( \alpha \), each \textit{v} is multiplied to its attention weight \( \alpha\textsubscript{i} \) and summed to obtain the attended output \( \hat{V\textsubscript{i}}\textsuperscript{att} \):

\vspace{-3mm}
\begin{gather}
s_i = tanh(M\textsuperscript{T}v_i + b_i) 
\\
\alpha_i = \frac{exp(s\textsubscript{i}\textsuperscript{T}c)}{\sum_{i=1}^{W}exp(s\textsubscript{i}\textsuperscript{T}c)} 
\\
\hat{V\textsubscript{i}}\textsuperscript{att} = \sum_{i=1}^{W} \alpha\textsubscript{i} v\textsubscript{i}
\end{gather}
\vspace{-3mm}

This is equivalent of learning how much previous information to attend when modeling relative changes in user behaviors by learning the weight of each turn.

\vspace{-1mm}
\paragraph{\bf Fully Connected Layer}
The outputs from contextualized word encoders, char encoders and attended feature matrix are concatenated to obtain each turn representation:

\vspace{-3mm}
\begin{gather}
Turn\textsubscript{i} = [Encoder\textsubscript{word}; \ Encoder\textsubscript{char}; \ \hat{V}\textsubscript{att}]
\end{gather}
\vspace{-3mm}

To benefit from all previous turn outputs, we have one final unidirectional LSTM that models each turn sequentially. Depending on tasks (online or offline prediction), many-to-many or many-to-one output(s) can be obtained. Each output is fed to a linear layer with dropout to enforce regularization, followed by sigmoid or softmax activation to obtain binary or multi-class distribution.

\vspace{-1.5mm}
\subsection{Behavioral Features}
Behavioral features extracted for ConvSAT are categorized into three types: 1) general behavioral features; 2) system features; 3) topic preference features. These features are concatenated to produce one feature vector per each turn.

\vspace{-1mm}
\paragraph{\bf General Behavioral Features}
General behavioral features are features that encode user behaviors in various dimensions, including lexical, semantics and conversational. First, we define engagements as subsets of conversation that have 4+ conversational depth on the same topic. Count of engagements (\textit{F}\textsubscript{1}) and max length of engagements (\textit{F}\textsubscript{2}) are derived respectively. Sentiment analysis using Valence Aware Dictionary for sEntiment Reasoning (VADER) \citep{vader} on utterances is applied to obtain positive (\textit{F}\textsubscript{3}, \textit{F}\textsubscript{5}) and negative (\textit{F}\textsubscript{4}, \textit{F}\textsubscript{6}) sentiment scores. To capture how much topic transition occurs, state change ratio (\textit{F}\textsubscript{7}) is derived by dividing total transitions to the current turn index. Similarly, agreement and disagreement ratios are derived (\textit{F}\textsubscript{8}, \textit{F}\textsubscript{9}) based on intent classification results. To measure the repetition between (\textit{U\textsubscript{i}}, \textit{R\textsubscript{i}}), (\textit{R\textsubscript{i-1}}, \textit{R\textsubscript{i}}) and (\textit{U\textsubscript{i-1}}, \textit{U\textsubscript{i}}), counts of token overlaps are computed (\textit{F}\textsubscript{10}, \textit{F}\textsubscript{11}, \textit{F}\textsubscript{12}). Lastly, the average and total word count of user utterances and system responses are extracted (\textit{F}\textsubscript{13} ... \textit{F}\textsubscript{18}). 

\vspace{-3mm}
\begin{center}
\footnotesize
\begin{table}[h]
    \begin{tabular}{l|l}
    \toprule
    \bf Local Features & \bf Short Description \\
    \bottomrule
        \textit{F}\textsubscript{1} - \textit{NumEngagements} &
        \#Engagements \\
        
        \textit{F}\textsubscript{2} - \textit{MaxEngagements} &
        Max engagement in \# of turns \\
        
        \textit{F}\textsubscript{3} - \textit{UtterancePos} &
        Positive sentiment in \textit{U\textsubscript{i}} \\
        
        \textit{F}\textsubscript{4} - \textit{UtteranceNeg} &
        Negative sentiment in \textit{U\textsubscript{i}} \\
        
        \textit{F}\textsubscript{5} - \textit{AvgPos} &
        Sum of pos sentiment counts / \textit{i} \\
        
        \textit{F}\textsubscript{6} - \textit{AvgNeg} &
        Sum of neg sentiment counts / \textit{i} \\
        
        \textit{F}\textsubscript{7} - \textit{StateChangeRatio} &
        \#Topic Transitions / \textit{i} \\
        
        \textit{F}\textsubscript{8} - \textit{YesRatio} & 
        \#Yes Responses/Agreements / \textit{i} \\
        
        \textit{F}\textsubscript{9} - \textit{NoRatio} & 
        \#No Responses/Disagreements / \textit{i} \\
        
        \textit{F}\textsubscript{10} - \textit{TokenOverlap\textsubscript{U}} &
        Token overlap in \textit{U\textsubscript{i}, U\textsubscript{i-1}} \\
        
        \textit{F}\textsubscript{11} - \textit{TokenOverlap\textsubscript{R}} &
        Token overlap in \textit{R\textsubscript{i}, R\textsubscript{i-1}} \\
        
        \textit{F}\textsubscript{12} - \textit{TokenOverlap\textsubscript{UR}} &
        Token overlap in \textit{U\textsubscript{i}, R\textsubscript{i}} \\
        
        \textit{F}\textsubscript{13} - \textit{TotalWord\textsubscript{U}} &
        Total \#Words in \textit{U\textsubscript{i}} \\
        
        \textit{F}\textsubscript{14} - \textit{TotalWord\textsubscript{R}} &
        Total \#Words in \textit{R\textsubscript{i}} \\
        
        \textit{F}\textsubscript{15} - \textit{AvgWord\textsubscript{U}} &
        Average \#Words in \textit{U\textsubscript{1}} ... \textit{U\textsubscript{i}} \\
        
        \textit{F}\textsubscript{16} - \textit{AvgWord\textsubscript{R}} &
        Average \#Words in \textit{R\textsubscript{1}} ... \textit{R\textsubscript{i}} \\ 
        
        \textit{F}\textsubscript{17} - \textit{Word\textsubscript{U}} &
        \#Words only in \textit{U\textsubscript{i}} \\
        
        \textit{F}\textsubscript{18} - \textit{Word\textsubscript{R}} &
        \#Words only in \textit{R\textsubscript{i}} \\
        
    \bottomrule
    \end{tabular}
    \end{table}
\end{center}
\vspace{-8mm}

\vspace{-1mm}
\paragraph{\bf System Features}
System features are directly related to systematic aspects of our conversational agent. There are two binary session-level features that capture if a user agreed to provide his name or if he is a returning user (\textit{F}\textsubscript{19}, \textit{F}\textsubscript{20}). For latency, we define two types, which are system latency (\textit{F}\textsubscript{21}, \textit{F}\textsubscript{22}, \textit{F}\textsubscript{23}) and user latency (\textit{F}\textsubscript{24}, \textit{F}\textsubscript{25}, \textit{F}\textsubscript{26}), both measured in seconds. System latency measures how long a user had to wait to hear the system response; user latency measures how long a user had to think before issuing an utterance. Lastly, every token in our utterances was annotated with ASR confidence value ranging from 0.0 to 1.0. Using these values, minimum, maximum and average token confidence on each \textit{U\textsubscript{i}} are added  (\textit{F}\textsubscript{27}, \textit{F}\textsubscript{28}, \textit{F}\textsubscript{29}).

\vspace{-3mm}
\begin{center}
\footnotesize
    \begin{table}[h]
    \begin{tabular}{l|l}
    \toprule
    \bf Session-level Features & \bf Short Description \\
    \bottomrule
        \textit{F}\textsubscript{19} - \textit{NameProvided} & Name provided or not \\
        \textit{F}\textsubscript{20} - \textit{ReturningUser} & Returning user or not \\
    \toprule
    \bf Local Features & \bf Short Description \\
    \bottomrule
        \textit{F}\textsubscript{21} - \textit{Latency} &
        System latency on \textit{U\textsubscript{i}} \\
        
        \textit{F}\textsubscript{22} - \textit{Latency\textsubscript{avg}} &
        Average system latency \\
        
        \textit{F}\textsubscript{23} - \textit{Latency\textsubscript{max}} &
        Max system latency \\
        
        \textit{F}\textsubscript{24} - \textit{UserLatency} &
        User latency on \textit{R\textsubscript{i}} \\
        
        \textit{F}\textsubscript{25} - \textit{UserLatency\textsubscript{avg}} &
        Average user latency \\
        
        \textit{F}\textsubscript{26} - \textit{UserLatency\textsubscript{max}} &
        Max user latency \\ 
        
        \textit{F}\textsubscript{27} - \textit{ASR\textsubscript{min}} & Min token confidence on \textit{U\textsubscript{i}} \\
        
        \textit{F}\textsubscript{28} - \textit{ASR\textsubscript{max}} & Max token confidence on \textit{U\textsubscript{i}} \\
        
        \textit{F}\textsubscript{29} - \textit{ASR\textsubscript{avg}} & Average token confidence on \textit{U\textsubscript{i}} \\
        
    \bottomrule
    \end{tabular}
    \end{table}
\end{center}
\vspace{-8mm}

\vspace{-1mm}
\paragraph{\bf Topic Preference Features}
Topic distribution features encode specific behaviors related to topic diversity, visited topics and topic distribution so far. For topic diversity, we counted the length of the visited topic set to represent topic breadth (\textit{F}\textsubscript{30}). Count of accepted topics and rejected topics (\textit{F}\textsubscript{31}, \textit{F}\textsubscript{32}) are extracted to explore topic acceptance and rejection trade-offs. Lastly, a 15-dim topic count vector and a 3-dim special state count vector from \textit{T\textsubscript{0}} to \textit{T\textsubscript{i}} are concatenated to represent the online topic distribution (\textit{F}\textsubscript{33}, ... \textit{F}\textsubscript{51}). The special states include \underline{Stop}, \underline{Profanity} and \underline{Clarification}. Stop state tracks whether a user expressed stop signals, profanity state tracks if an utterance or response contained profane words, clarification state tracks if system asked a user to repeat due to low ASR confidence.

\begin{center}
\footnotesize
    \vspace{-1mm}
    \begin{table}[h]
    \begin{tabular}{l|l}
    \toprule
    \bf Local Features & \bf Short Description \\
    \bottomrule
        \textit{F}\textsubscript{30} - \textit{TopicBreadth} & Number of unique topics visited \\
        \textit{F}\textsubscript{31} - \textit{TotalAcceptedTopics} & \#Accepted topics \\
        \textit{F}\textsubscript{32} - \textit{TotalRejectedTopics} & \#Rejected topics \\
        \textit{F}\textsubscript{33...51} - \textit{TopicDistribution} & Vector of 18 topic counts \\
    \bottomrule
    \end{tabular}
    \end{table}
\end{center}
\vspace{-13mm}


\vspace{-1.5mm}
\subsection{Additional Implementation Details}
For contextualized word encoders, embedding weights are initialized with pretrained Google Word2Vec \citep{mikolov2013distributed} of size 300 and tuned for conversational context. For contextualized char encoders, embedding weights of size 32 are randomly initialized and learned during training. We used 3 for \textit{W}, since we observed adding less or more context reduced performance on our experiments. Hidden dimension size 100 is used for each word LSTM and 32 for each char LSTM, resulting in each turn representation of size 528 (utterance + response) + \#features. Adam optimizer was used to minimize cross entropy loss, with a 1e-4 learning rate. At the fully connected layer, a dropout rate of 0.5 is used. These hyper-parameters were obtained after tuning them to our Alexa validation data, but can be easily tuned for different conversational tasks. Our PyTorch implementation and models are available for the research community\footnote{Available at {\em {\url{https://github.com/emory-irlab/ConvSAT}}}}.

\section{Conversational Data and Tasks}
We present statistics of DBDC3 dataset and our private dataset collected during Amazon Alexa Prize 2018. Then, three classification tasks are defined based on these two datasets.

\vspace{-1.5mm}
\subsection{Dialogue Breakdown Detection Challenge}
Dialogue system technology challenges (DSTC), originally known as the dialogue state tracking challenges, were initiated in 2013 in order to promote research in conversational AI. We focus on the third track of DSTC6'17 challenge titled Dialogue Breakdown Detection Challenge 3 (DBDC3) \citep{dbdc3}, since it is closely related to online satisfaction prediction. Dialogue breakdown is defined as a situation in conversations where users cannot continue engaging with the system due to various system failures. Table \ref{data_stats} summarizes the DBDC3 English corpus statistics.

\vspace{-2mm}
\begin{table}[ht]
      \centering
          \begin{tabular}{l|l|l|l}
           \toprule
            & \textbf{Training} & \textbf{Val} & \textbf{Test}  \\
            \bottomrule
            \textbf{Dialogues} & 373 & 42 & 200 \\
            \textbf{Turns} & 3730 & 420 & 2000 \\
            \textbf{NB} & 1207 (32.3\%) & 126 (33.3\%) & 756 (37.8\%) \\
            \textbf{PB} & 974 (26.1\%) & 114 (27.1\%) & 456 (22.8\%) \\
            \textbf{B} & 1549 (41.5\%) & 180 (42.8\%) & 788 (39.4\%) \\
          \bottomrule
         \end{tabular}
      \caption{Dialogue Breakdown Detection Challenge 3 data statistics (English corpus). ``NB'' stands for not breakdown, ``PB'' stands for potential breakdown, ``B'' stands for breakdown.}
      \label{data_stats}
\end{table} 
\vspace{-7mm}

Each turn is labeled by 30 human annotators with three labels: 1) not breakdown (NB); 2) potential breakdown (PB); 3) breakdown (B). According to the task specification, turn labels are obtained from majority voting and have to be predicted without looking at future context. We use the official training and test data splits to be consistent with other models published on this data. For our model training, we further set aside 10\% of the official training data for model validation.

\vspace{-1.5mm}
\subsection{Alexa Prize Dataset}
\paragraph{\bf Alexa Prize Data Overview}

Alexa Prize Dataset was collected during a worldwide research competition sponsored by Amazon, initiated in 2017 to advance conversational AI \citep{conver} and continued in 2018. Our system conversed with thousands of Alexa customers during summer 2018, providing the ``Alexa Prize'' dataset for this paper. Customers were invited to optionally provide a rating when they were finished talking to the bot. Rated dialogues received rating scores between 1.0 and 5.0. A small subset (less than 1\%) of the users who rated our system also provided free-form feedback, explaining why they chose their rating. For this study, we will only focus on conversations from one stable version of our system, with the data collected over a 2-week period in August 2018. The data used for this study contained 5,044 rated conversations, with 4,811 conversations (95.3\%) from unique users. We randomly selected 93 conversations as our test set, and selected an additional 10\% of the remainder as our validation set for training. Table \ref{alexa_stats} reports the statistics for Training, Validation, and Test data splits. 
\vspace{-1mm}
\begin{table}[h]
    \centering
    \begin{tabular}{l|l|l|l}
    \toprule
    & \textbf{Training} & \textbf{Val} & \textbf{Test}  \\
    \bottomrule
    \textbf{Dialogues} & 4455 & 496 & 93 \\
    \textbf{Turns} & 80996 & 8864 & 1959 \\
    \textbf{Turns\textsubscript{avg}} & 18.18 & 17.87 & 21.06 \\
    \textbf{Rating\textsubscript{1}} & 593 (13.3\%) & 62 (12.5\%) & 10 (10.7\%) \\
    \textbf{Rating\textsubscript{2}} & 671 (15.0\%) & 74 (14.9\%) & 11 (11.8\%) \\
    \textbf{Rating\textsubscript{3}} & 811 (18.2\%) & 95 (19.1\%) & 17 (18.2\%) \\
    \textbf{Rating\textsubscript{4}} & 860 (19.3\%) & 96 (19.3\%) & 19 (20.4\%) \\
    \textbf{Rating\textsubscript{5}} & 1520 (34.1\%) & 169 (34.0\%) & 36 (38.7\%) \\
    \bottomrule
    \end{tabular}
    \caption{Alexa Prize 2018 data statistics.}
    \label{alexa_stats}
\vspace{-5mm}
\end{table} 
\vspace{-2mm}

For the entire data, the standard deviation on turns is \textbf{15.81}, meaning our data covers a wide range of different conversations from extremely short, to very long ones, with some conversations lasting over 100 turns. Interestingly, there was no strong correlation between a user rating and conversation length: the Pearson correlation coefficient is \textbf{0.095}, indicating no correlation. Lastly, our system supports conversations on 15 different domains, ranging from popular domains such as \textit{Movies} and \textit{Music} to generic domains such as \textit{Weather} and \textit{Wikipedia}. Our domain classifier, described in reference \cite{ConCET} achieved \textbf{0.717} Micro-Averaged F1 on our 3,000 annotated test utterances.

\begin{figure}[h]
\centering
  \includegraphics[width=0.7\columnwidth]{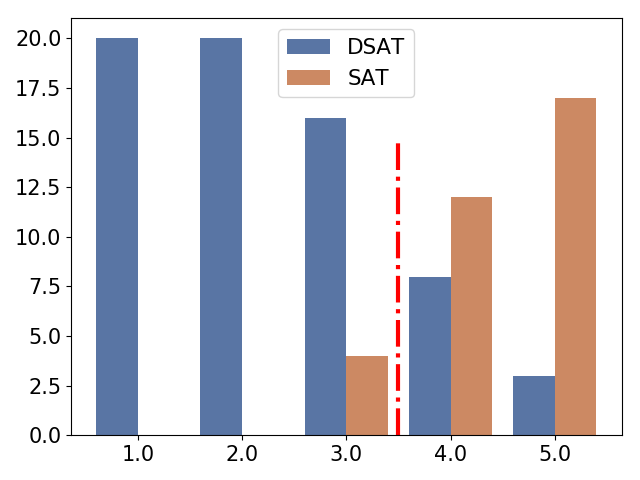}
\vspace{-5mm}
\captionof{figure}{Count (y-axis) of dissatisfied and satisfied feedback among different rating groups (x-axis). The red line indicates the best cut (rating=3.5) between SAT/DSAT labels.}
\label{feedback}
\end{figure}

\vspace{-1mm}
\paragraph{\bf User rating vs user satisfaction}
User rating and user satisfaction are clearly related, but they are different metrics. In non-goal oriented setting, user rating is very subjective and hard to generalize especially in the five groups defined in Table \ref{alexa_stats}. To simplify and better generalize our study, we propose to find a statistical relationship between user rating and user satisfaction using user feedback. We randomly selected 20 free-form feedback each from five rating groups and asked one human annotator to label each feedback as satisfied or dissatisfied. The goal is to find a threshold that best splits satisfaction (SAT) and dissatisfaction (DSAT). Our annotation results are reported in Figure \ref{feedback}. 

For the experiments in this paper, we chose to frame the problem as a binary classification task, to predict SAT (satisfaction) vs. DSAT (dis-satisfaction). There is a long tradition in evaluation literature for this approach, e.g., \citep{kiseleva2016understanding, hancock2019learning, pred, egregious, sensitive} in order to reduce high subjectivity and noise in user ratings. The challenge is where to choose the boundary to convert the user ratings to SAT/DSAT decisions. To set the DSAT/SAT boundary, we performed a qualitative analysis of user feedback. The qualitative results indicate that for 1.0 and 2.0 rating groups, 100\% of users left negative feedback based on their interactions. For the 3.0 rating group, we see a small increase in positive feedback, but still, 80\% of users were dissatisfied. For 4.0 and 5.0 rating groups, only 40\% and 15\% of users were dissatisfied. Hence, we conclude that setting a boundary between 3.0 and 4.0 ratings will best separate dissatisfaction from satisfaction, and we define our two user satisfaction labels as DSAT (ratings <= 3.5) and SAT (ratings > 3.5). Defining SAT to correspond to ratings of over 3.5 out of 5 has an additional benefit. One important goal of online satisfaction prediction is to provide consistent and reliable reinforcement signals for tasks such as online dialogue policy learning or model tuning. For such tasks, knowing highly satisfactory (and strongly dis-satisfactory) outcomes is valuable, while intermediate ``partially'' satisfied signals are not helpful.

\vspace{-3mm}
\vspace{-5mm}

\paragraph{\bf Annotating online satisfaction labels}
We defined SAT and DSAT labels based on our user feedback analysis. However, user ratings were requested {\em after} the conversation ended, and do not provide online satisfaction labels. To solve this challenge, we had to annotate each turn in a consistent and reliable way. To obtain these ground truth labels, we asked two human annotators to label our 1,959 turns using the annotation guidelines below. Only the conversation transcripts data (utterances and responses) were provided during the annotation process.

\begin{itemize}
    \item \textit{Label each turn into SAT or DSAT by considering all the previous information up to the current turn.}
    \item \textit{Factors to consider are conversational depth within the current topic, conversational coherency, domain detection rate, response quality, topic diversity, ASR and other miscellaneous errors.}
\end{itemize}



For offline predictions, we use the satisfaction label derived from real ratings. Hence, the number of offline samples (93) is identical to the number of dialogues (93). For online predictions, we predict on all previous turns except for the final turn, resulting 1866 (1959-93) samples. The final SAT class distribution of offline and online test samples are 40.9\% and 56.8\% respectively. The kappa score \citep{kappa} between the two annotators on these 1866 samples is \textbf{0.753}, showing a substantial agreement. In the case of a disagreement, the final label was randomly chosen. 

\vspace{-2mm}
\subsection{Task Definition}
Based on the datasets, we define three classification tasks: 1) dialogue breakdown detection; 2) online satisfaction prediction; 3) offline satisfaction prediction.


\paragraph{\bf Dialogue Breakdown Detection}
\vspace{-1mm}

\label{dbc_task}
Given a conversation turn (\textit{i}), which is a concatenated vector of [\textit{U\textsubscript{i}\textsuperscript{w}}; \textit{R\textsubscript{i}\textsuperscript{w}}; \textit{U\textsubscript{i}\textsuperscript{c}}; \textit{R\textsubscript{i}\textsuperscript{c}};
\textit{\( \hat{V\textsubscript{i}} \)}] defined in Section \ref{approach}, predict the dialogue breakdown label \textit{B\textsuperscript{i}\textsubscript{pred}} of each turn:

\vspace{-5mm}
\begin{gather}
B\textsuperscript{i}_{pred} \in{(NB, PB, B)} 
\end{gather}
where {\em NB}, {\em PB}, and {\em B} represent ``not breakdown'', ``possible breakdown'' and ``breakdown'', respectively.

\label{satisfaction_task}

\paragraph{\bf Online Satisfaction Prediction}
\vspace{-1mm}

We define two states for the dialogue: {\em DSAT} for dis-satisfied (equivalent to ``breakdown'') and {\em SAT} for satisfied (equivalent to ``not breakdown''). Given each \textit{T\textsubscript{i}}, conditioned on previous turns, we predict the most likely binary satisfaction label \textit{S\textsuperscript{i}\textsubscript{pred}} of each turn:

\vspace{-3mm}
\begin{gather}
S^i_{pred} \in{(SAT, DSAT)}
\end{gather}

\paragraph{\bf Offline Satisfaction Prediction}
\vspace{-1mm}

Given a session of length \textit{N} turns, we predict \( S^{N}_{pred} \) at the end (\textit{T\textsubscript{N}}) of the conversation:

\vspace{-3mm}
\begin{gather}
S^{N}_{pred} \in{(SAT, DSAT)} 
\end{gather}
Note that at the last turn of the conversation, the online- and offline- prediction tasks are equivalent.

\section{Experimental Setup}
For Alexa data, one remaining challenge is to create large-scale training samples for online prediction since human labeling is expensive and applied only to the test set. We introduce the labeling functions we devised to heuristically create large-scale training samples. Then, an overview of baseline models and evaluation metrics is presented.

\paragraph{\bf Data Programming for Alexa Training Data}
\vspace{-1mm}

Since online satisfaction annotation is extremely time-consuming, it is not feasible to generate all the necessary labels for training. Moreover, because of privacy issues with Amazon customers, we cannot outsource the annotation task to a public service like Amazon Mechanical Turk. Given the small size of human-labeled data, training on it is unrealistic. Based on these limitations, our proposed solution is to apply data programming to generate training data by using heuristic weak supervision strategies. We combine our domain heuristics to design a set of simple rule-based labeling functions \citep{data_pro, egregious} to generate online training labels. Once large-scale training data is generated, the goal is to compare heuristic performance with proposed models to see if models can learn beyond these simple rules. The details of our labeling process are described below.

\begin{itemize}
    \item \textit{Label SAT for each engagement of depth >= 4}
    \item \textit{Label SAT for 4+ consecutive affirmation intents}
    \item \textit{Label DSAT for 4+ consecutive negation intents}
    \item \textit{For remaining unlabeled turns, use imputation}
\end{itemize}

For Alexa data, the average engagement depth on various domains is \textbf{2.45}, ranging from \textbf{3.20} for the most popular \textit{"Movies"} domain and \textbf{2.11} for the least popular \textit{"Travel"} domain. Hence. we heuristically define \textbf{4+} engagement depth as a successful signal. Affirmation intents indicate users agreeing to the system's recommendations while negation intents indicate disagreement and topic switches, which provide intuitions for the second and third rules. Lastly, the remaining unlabeled turns are imputed based on the average ratings from beginning to the current turn \textit{T\textsubscript{i}}, similar to our proposed online scaling mechanism in Section 3.3. We consider SAT labels as 5.0 ratings and DSAT labels as 1.0 ratings during imputation. The last turn \textit{T\textsubscript{n}} label always follows the real user rating.

To measure the statistical correlation of heuristic labeling to human annotated baseline, we applied these rules to our test data and computed the Fleiss Kappa score. The Kappa score is \textbf{0.46}, indicating moderate agreement. Hence, we hypothesize that these rules are reliable heuristics to generate large-scale training data. We emphasize that the heuristic labeling was done to generate training data only. The test data was manually annotated by two independent internal judges.

\paragraph{\bf Methods Compared}
\vspace{-1mm}

We define our first baseline method as a non-contextual bi-LSTM model (LSTM). This model only looks at the current utterance and response, which is equivalent of setting contextual window size \textit{W} as 1. For state-of-the-art (SOTA) baseline, a contextual bi-LSTM (CLSTM), introduced by Hashemi et al. \citep{sensitive}, models word-level contextual utterance representations along with conversational history. For DBDC3 data, we additionally report the best performing model (KTH Entry) participant on this challenge, which is a non-contextual LSTM  model combined with a bag-of-words, Doc2Vec embeddings, and manual features \citep{kth}. Additionally, heuristic labeling (HL) baseline is reported for the online satisfaction task.


\paragraph{\bf Evaluation Metrics}
\vspace{-1mm}

For DBDC3 task, we stay consistent with the official evaluation metrics, which are Micro-Averaged Accuracy and Macro-Averaged F1 on the breakdown label. We will additionally report Precision and Recall on breakdown labels for our implemented models. For the Alexa dataset, consistent with the DBDC3 setup, we report the Micro-Averaged Accuracy and Macro-Averaged values of Precision, Recall, and F1 scores for both SAT and DSAT classes. 

\paragraph{\bf Training Details}
\vspace{-1mm}

For DBDC3 data, since our behavioral features are designed for our Alexa Prize system, some of the features related to latency, ASR, and detailed topic-specific features are not available. Hence, these features are excluded when training on DBDC3 data. For word encoders, the hidden dimension was set to 64 to prevent overfitting. We used softmax activation on output layers for DBDC3 data (since it is a multi-class problem) and sigmoid activation for Alexa data (more appropriate for the binary classification problem). All the other settings, including the model architecture (described in Section 3.3) remained identical.

\section{Main Results}
In this section, we compare ConvSAT to other baselines on three tasks defined in Section \ref{satisfaction_task}.

\paragraph{\bf Dialogue Breakdown Detection Results}
\vspace{-1mm}

ConvSAT significantly outperformed all the baseline models on Accuracy, Precision, Recall and F1 for dialogue breakdown detection task, as shown in Table \ref{dbc_results}. There are 14.7\% and 36.1\% improvement in Accuracy and F1 compared to KTH entry. Precision and Recall for KTH entry are left blank because the official metrics did not include these. Similarly, ConvSAT improved the SOTA baseline by 2.4\% on Accuracy and 5.5\% on F1 score, indicating statistically significant improvements with $p<0.05$, measured by two-tailed Student's t-test. To ensure stability of the results and improvements, we report the mean and standard deviation of ConvSAT performance on five random test folds of 40 conversations each. Higher deviations in Recall mostly occur between B and PB labels, indicating that the distinction between these two labels are the most challenging. Nonetheless, it is clear that leveraging sub-word information and behavioral feature matrices are beneficial for predicting failure. 


\vspace{-1mm}
\begin{table}[h]
\footnotesize
      \centering
          \begin{tabular}{l|c|c|c|c}
           \toprule
            \textbf{Model} & \textbf{AC} & \textbf{PR(B)} & \textbf{RC(B)} & \textbf{F1(B)} \\
            \bottomrule
            KTH Entry & 0.441 & - & - & 0.349\\
            LSTM  & 0.456 & 0.322 & 0.566 & 0.410\\
            CLSTM  & 0.494 & 0.351 & 0.625 & 0.450\\
            \midrule
            \textbf{ConvSAT} & \textbf{0.506*}$\pm$0.9 & \textbf{0.374*}$\pm$0.8
            & \textbf{0.651*}$\pm$2.6  & \textbf{0.475*}$\pm$1.0 \\
             Impr. over KTH  & 14.7\% & - & - & 36.1\% \\
             Impr. over LSTM  & 10.9\% & 16.1\% & 15.0\% & 15.8\% \\
             Impr. over CLSTM  & 2.4\% & 6.5\% & 4.1\% & 5.5\% \\
          \bottomrule
         \end{tabular}
      \caption{Accuracy (AC), Precision (PR), Recall (RC) and F1 scores for dialogue breakdown detection. ``B'' stands for the breakdown label. ``*'' indicates statistical significance of improvement based on two-tailed Student's t-test with $p<0.05$, compared to CLSTM.}
    \label{dbc_results}
    \vspace{-5mm}
\end{table} 
\vspace{-1mm}

We highlight that there is a significant gap in KTH entry and our re-implemented LSTM baseline (the LSTM baseline exhibits higher performance). The reason is due to a seemingly minor change in utterance representation. For KTH entry in the DBDC3 challenge, each utterance was represented by averaging the Google's Word2Vec embeddings with pre-trained vectors, while our implementation of the LSTM baseline considers each word separately. This is significant because averaging simplifies the training process but loses the temporal relationship between each word. Moreover, KTH entry represented each turn differently from our LSTM baseline by treating each utterance and response as separate timestamps. This doubles the length of the original sequence, and required insertion of dummy labels for each utterance to satisfy the length of predictions to be same as the input. During prediction, the $argmax$ on three true labels were applied to each system response, ignoring the dummy label. In contrast, our LSTM baseline avoids this complexity by having two separate networks to represent each utterance and response separately. As a result, since our re-implementation of the baseline LSTM-based approach (inspired by the KTH entry) exhibits substantially higher performance on all metrics on this benchmark dataset, we use our LSTM implementation as the baseline for all subsequent Alexa experiments.

\paragraph{\bf Online Satisfaction Prediction Results}
\vspace{-1mm}

ConvSAT improved all three baseline models on the online satisfaction prediction task, as reported in Table \ref{real-time-result}, with significant improvements over all the baselines on all metrics. This provides strong evidence that behavioral features and character information enable significant gains in real-world conversations. Compared to our heuristic baseline, ConvSAT showed 7.8\% improvement in both Accuracy and F1 respectively. Compared to the recent SOTA baseline, ConvSAT also improved by 2.4\% and 2.2\% on Accuracy and F1 respectively, with all improvements significant with $p<0.05$.


\begin{table}[h]
\footnotesize
      \centering
          \begin{tabular}{l|l|l|l|l}
           \toprule
            \textbf{Model} & \textbf{AC} & \textbf{PR} & \textbf{RC} & \textbf{F1} \\
            \bottomrule
            HL  & 0.735 & 0.731 & 0.728 & 0.729\\
            LSTM  & 0.749 & 0.763 & 0.732 & 0.734 \\
            CLSTM  & 0.774 & 0.772 & 0.767 & 0.769\\
            \midrule
            \textbf{ConvSAT} & \textbf{0.793*}$\pm$0.8 & \textbf{0.795*}$\pm$1.6
            & \textbf{0.783*}$\pm$1.4  & \textbf{0.786*}$\pm$1.3 \\
             Impr. over HL  & +7.8\% & +8.7\% & +7.5\% & +7.8\% \\
             Impr. over LSTM  & +5.8\% & +4.1\% & +6.9\% & +7.0\% \\
             Impr. over CLSTM  & +2.4\% & +2.9\% & +2.0\% & +2.2\% \\
          \bottomrule
         \end{tabular}
      \caption{Online satisfaction prediction Accuracy, Precision, Recall and F1 scores for detecting the SAT label in the Alexa Prize 2018 dataset.}
      \label{real-time-result}
      \vspace{-5mm}
\end{table} 


ConvSAT achieved 0.786 Precision, 0.865 Recall and 0.823 F1 for the DSAT label. For the SAT label, 0.804 Precision, 0.701 Recall and 0.749 F1 were achieved. The standard deviations are also computed based on random 5 test folds. This shows that predicting SAT label correctly is harder than correctly classifying DSAT label. Intuitively, satisfactory conditions should be more subjective than failure conditions because people can still dislike the conversation simply because the responses are boring or lack coherence. However, there are more explicit signals of failures, such as low ASR confidence, profane utterances and high latency. 


\paragraph{\bf Offline Satisfaction Prediction Results}
\vspace{-1mm}

For offline satisfaction prediction, we noticed that the general performance is lower compared to the online prediction results. This is because offline satisfaction prediction requires more complex reasoning that spans from the beginning to the end of conversations. Since our conversations have, on average, over 16 turns, we expect the decision boundaries to be more complex.

\vspace{-1mm}
\begin{table}[h]
\footnotesize
      \centering
          \begin{tabular}{l|l|l|l|l}
           \toprule
            \textbf{Model} & \textbf{AC} & \textbf{PR} & \textbf{RC} & \textbf{F1} \\
            \bottomrule
            LSTM  & 0.656 & 0.679 & 0.683 & 0.656\\
            CLSTM  & 0.709 & 0.706 & 0.717 & 0.705\\
            \midrule
            \textbf{ConvSAT} & \textbf{0.731*}$\pm$2.1 & \textbf{0.738*}$\pm$0.7
            & \textbf{0.750*}$\pm$1.0  & \textbf{0.729*}$\pm$2.0 \\
            Impr. over LSTM  & 11.4\% & 8.6\% & 9.8\% & 11.1\% \\
            Impr. over CLSTM  & 3.1\% & 4.5\% & 4.6\% & 3.4\% \\
          \bottomrule
         \end{tabular}
      \caption{Offline satisfaction prediction Accuracy, Precision, Recall and F1 scores for detecting the SAT label in the Alexa Prize 2018 dataset.}
    \label{turn_results}
    \vspace{-5mm}
\end{table} 

Nonetheless, ConvSAT outperforms the two state of the art baseline models significantly. There are 11.4\%, 11.1\% increases in Accuracy and F1, respectively, compared to the non-contextual LSTM, and 3.1\%, 3.4\% boost compared to the contextual LSTM baseline. ConvSAT achieved 0.864 Precision, 0.667 Recall and 0.752 F1 for DSAT. For SAT labels, ConvSAT achieved 0.612 Precision, 0.833 Recall, and 0.706 F1 score, which follows a similar pattern to online satisfaction results.


\vspace{-1mm}
\section{Discussion and Error Analysis}
To understand the impact of different features groups, we conducted a feature ablation study on ConvSAT by systematically removing text representation and behavioral features. Then, we present the top 10 strongest behavioral features for Alexa dataset, followed by error analysis.

\paragraph{\bf Feature Ablation}
\vspace{-1mm}

To show the effect of behavioral features and character information, we conducted an ablation study on both datasets by systematically removing these portions from ConvSAT. Table \ref{ablation} shows the feature ablation results on online satisfaction and breakdown detection tasks. We used the same evaluation metrics defined for each task.

The results show that removing both behavioral features and character information decreases the Accuracy and F1 on both datasets. In general, the decrease is much greater when removing behavioral features over removing characters. It shows that word-level information already contains most information, and in the future, more advanced subword representation such as phonetic representation needs to be explored.

\begin{table}[h]
      \centering
          \begin{tabular}{r|l|l|l|l|l|l}
           \toprule
            \textbf{Model} & \textbf{AC(S)} & \textbf{F1(S)} & \textbf{AC(B)} & \textbf{F1(B)} & \textbf{BF} & \textbf{C} \\
            \bottomrule
            ConvSAT (full)  & 0.793 & 0.786 & 0.506 & 0.475 & \cmark & \cmark \\
            \midrule
            - Characters & 0.792 & 0.784 & 0.505 & 0.472 & \cmark & \xmark \\
            \%Change & -0.1\% & -0.2\% & -0.2\% & -0.6\% & - & - \\
            \midrule
            - Behavior & 0.773 & 0.769 & 0.494 & 0.450 & \xmark & \xmark \\
            \%Change  & -2.5\% & -2.1\% & -2.3\% & -5.2\% & - & - \\
          \bottomrule
         \end{tabular}
      \caption{Feature ablation on online satisfaction (S) and dialogue breakdown detection (B) tasks. ``BF'' and ``C'' stand for Behavioral features and Character features, respectively.}
    \label{ablation}
    \vspace{-5mm}
\end{table}

To conclude, distributional semantics are important features since they help models to learn the general context. However, we claim that they are not sufficient to model complex interactions between textual data and subjective satisfaction. For instance, a phrase \underline{I am done} can be a strong signal of dissatisfaction after recent failures. However, after several successful engagements on multiple topics, the same phrase can represent a satisfaction or topic completion signal. Using distributional semantics alone, the model is likely to generalize on more frequent cases without learning the conversational flow effectively. Hence, we conjecture that our model successfully captures the behavioral features' interaction with semantics, resulting in significant performance improvements over semantics alone.


\paragraph{\bf Importance of Behavioral Features}
\vspace{-1mm}

\begin{figure}[h]
\centering
  \includegraphics[width=200pt]{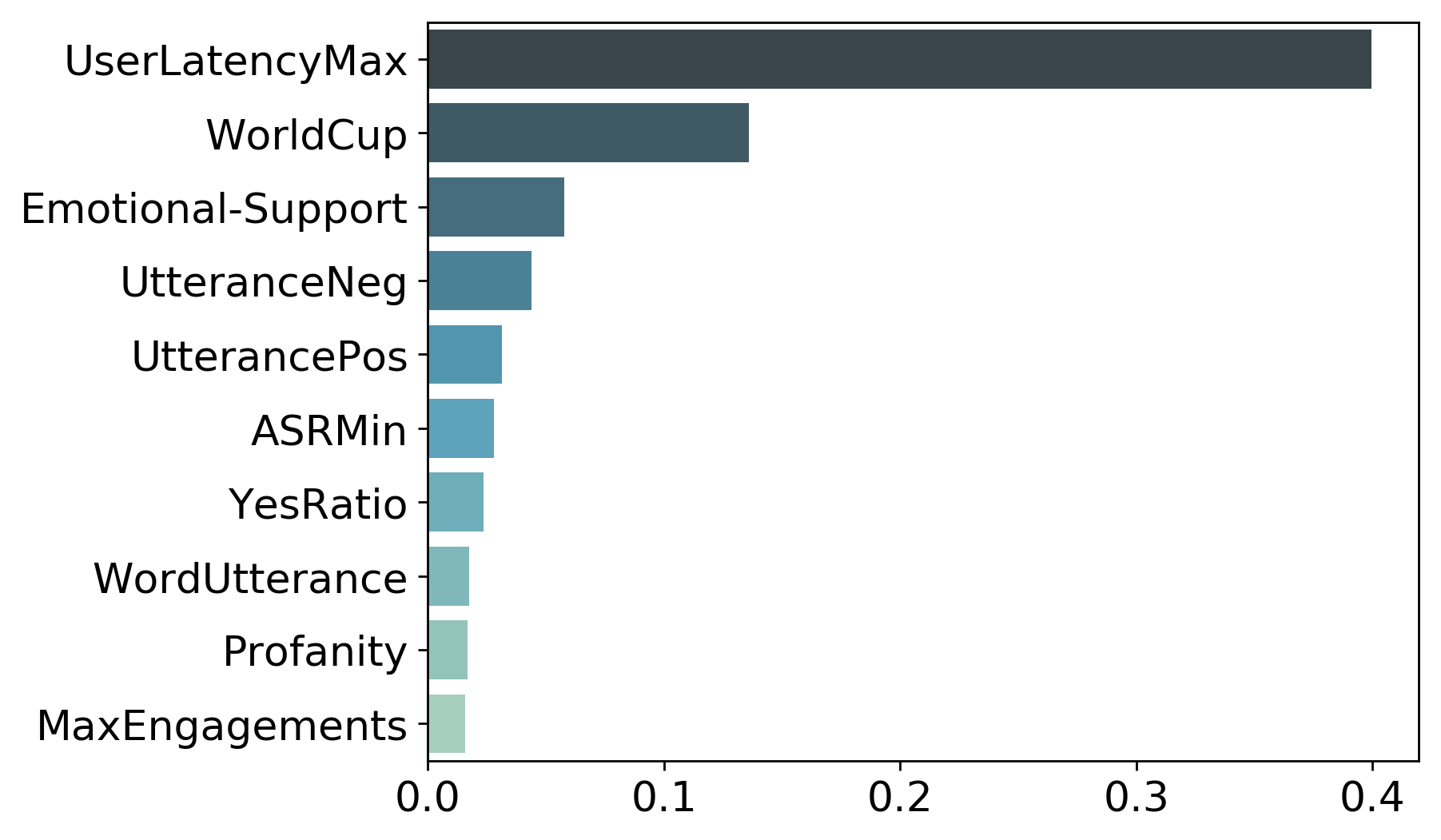}
\vspace{-3mm}
\captionof{figure}{Top 10 feature importance from gradient boosted decision tree, trained on online satisfaction Alexa data.}
\label{feature_importance}
\vspace{-2mm}
\end{figure}

Since we confirmed the importance of general behavioral signals, we now delve into specific behavioral feature importance. To understand the importance of each signal, we trained a gradient boosted decision tree (GBDT) by only using the behavioral feature matrices. We selected this tree-based model because of easy interpretability and support for categorical features. We used grid search to optimize the GBDT parameters, and used 5-fold cross validation to better generalize our model. Figure \ref{feature_importance} reports the top 10 features learned for this task, using binary logistic loss function. We trained GBDT only on online Alexa data because we have a more comprehensive set of features, and substantially larger samples compared to the DBDC3 dataset.

The results show that maximum user latency, UserLatencyMax (the amount of time a user spends before speaking) is a strong signal for online user (dis-)satisfaction. There are several potential reasons for high user latency, such as user not understanding the system's response, or simply does not know what to say after a nonsensical response. An increase in user latency also potentially compounds ASR errors, as the user's utterance is more likely to be cut off after a long pause.

Talking about the Worldcup was one of the most popular topics people enjoyed, while the emotional support topic was only invoked when user's mood seemed negative. These two features show that adding a topic distribution vector can help the model to learn the effect of an individual topic on user satisfaction. Sentiment features were also important because it is natural to expect a strong correlation between user sentiment and user satisfaction. Minimum ASR confidence value is another strong indicator of potential ASR failure and strongly contributed to importance. Lastly, utterance word counts, engagement (topic depth), and profanity (user cursing) are found to be strongly correlated with satisfaction and dis-satisfaction, respectively.

\paragraph{\bf Representative Error Analysis}
\vspace{-1mm}

Since we cannot release the real Alexa utterances, our error analysis is performed only on the DBDC3 data. After comparing the prediction results between ConvSAT and the SOTA baseline, we present a representative case study where both models failed. The true label is indicated inside the bracket next to the final response.



\vspace{2mm}
\begin{dialogue}
\footnotesize
\hrule
\speak{Bot} Hello, I really like movies. How about we talk about movies?
\speak{User} Sure. What's the most recent movie you've seen?
\speak{Bot} Honestly, you know, they're also different but we were talking before about the Wolf of Wall Street.
\speak{User} I liked that movie a lot. I also liked inception.
\speak{Bot} Well, you know, I'm not sure. But anyway, I'd like to chat with you. [Not Breakdown]
\vspace{1mm}
\hrule
\end{dialogue}
\vspace{2mm}

For this case, the CLSTM baseline predicted "breakdown" and ConvSAT predicted "potential breakdown". This is expected since there are many breakdown samples that contain phrases such as "I am not sure" because many bots simply avoid answering if they did not understand. However, for humans, this example is acceptable since the bot acknowledged its mistake and suggested to continue chatting. We believe that such human-like reasoning is challenging for neural networks even with a more advanced contextual representation. To be successful, improvements in representing and combining behavioral signals to context need more explorations.


\section{Conclusions and Future work}
Conversational agents are being used widely in information-search, online bookings, and almost any setting where a human interaction could be valuable. While much prior work focused on the implementation and science behind these agents, this paper focuses on developing new, automated ways to evaluate conversational agents in online using contextual and behavioral clues. 

We proposed a neural architecture called ConvSAT that combines these signals: 1) contextualized utterance and response representation; 2) contextualized sub-word information of utterance and response; 3) behavioral feature matrices; 4) previous conversational history. We experimented with thousands of real open-domain conversations as well as publicly available DBDC3 dataset to conduct a large-scale study on predicting satisfaction and dialogue breakdown. Our results are promising as ConvSAT outperformed state-of-the-art baselines in all three tasks, reaching 0.79 Accuracy and F1 on the SAT class for the online satisfaction prediction task. 

Our experiments demonstrate that aggregating multiple signals derived from user behavior, topic preferences, system state, distributional semantics, and conversational context is needed when designing a successful satisfaction prediction model. In addition, we presented insights derived from feature ablation and importance for these tasks, showing that latency, topical, sentiment and ASR features are strong predictors of user (dis-)satisfaction. To conclude, our new ConvSAT model of conversational satisfaction, and experiments in online satisfaction prediction, offer promise for adaptive conversation strategies. The predicted satisfaction could be used for both offline evaluation for improving conversational systems, or as online feedback for adapting the conversation for each user, enabling a new generation of more responsive and intelligent conversational agents.


\begin{acks}
We gratefully acknowledge the computational and technical support from Amazon Alexa Prize 2017 and 2018.
\end{acks}

\balance
\bibliographystyle{abbrv}
\bibliography{acmart}
\end{document}